\documentstyle[12pt,eqpersec]{article}

\begin{document}

\begin{center}
\large {\bf THE DIRAC-COULOMB PROBLEM FOR THE $\kappa$-POINCAR\'{E} QUANTUM
GROUP}

\vspace{46pt}
\normalsize L.C. Biedenharn \\ {\it Center for Particle Physics,Department of
Physics, University of Texas at Austin, \\ Austin, TX 78712}

\vspace{18pt}

Berndt Mueller \\ {\it Department of Physics, Duke University, Durham
NC, 27708}

\vspace{18pt}

Marco Tarlini \\ {\it INFN, Sezione di Firenze, Dipartimento di
Fisica, Universit\'{a} degli Studi di Firenze, \\ Largo E. Fermi 2, 50125
Firenze, Italy}

\vspace{36pt}
{\bf Abstract}

\end{center}

{\footnotesize The recently introduced $\kappa$-Poincar\'{e}-Dirac equation is
gauged to treat the $\kappa$-Dirac-Coulomb problem.  For the resulting
equation, we prove that the perturbation to first order in the quantum group
parameter vanishes identically.  The second order perturbation is singular, but
assuming a heuristic cut-off allows a qualitative estimate of the quantum
group parameter.}

\baselineskip=18pt

\section{Introduction}

\indent \indent Quantum groups are a new symmetry structure recently introduced
into physics.$^{1-3}$   For {\it semi-simple} Lie algebras,
the most important properties of the representation theory have been
generalized to their quantum analogs$^{1-6}$ and the theory and
applications are by now well-developed.  The situation is very different for
{\it non-semisimple} groups where the quantum deformation structure is not
unique.$^{7}$  Some of the most fundamental, and interesting, symmetries in
physics -- most importantly the Poincar\'{e} and Galilei
symmetries -- fall into this category.

One approach to a deformed Poincar\'{e} algebra is to apply the
contraction process to the standard q-deformation of the (anti) de
Sitter algebra $so(3,2)$.  Taking the limit of the de Sitter radius
$R \rightarrow \infty$ with
an accompanying limit of the deformation parameter $q$, such that
$ \lim (R \; \ln q) = \kappa^{-1}$, one obtains the {\it $\kappa$-Poincar\'{e}
quantum group}.$^{8}$

The $\kappa$-Poincar\'{e} quantum group has two invariants, a quadratic
invariant and a bi-quadratic invariant.  The $\kappa$-Dirac equation has
recently been found,$^{9}$ and it factorizes the bi-quadratic invariant, quite
unlike the usual case which factorizes the quadratic invariant.

To obtain the $\kappa$-Dirac-Coulomb equation we have gauged the $\kappa$-Dirac
equation and incorporated the Coulomb field.  Expanding this equation in the
(dimensional) parameter $\kappa^{-1}$, we apply it to the quantum
relativistic hydrogen atom, with the aim of confronting any induced
$\kappa$-Poincar\'{e} shift in the ground state energy level with the recent
precision quantum optics measurements.  Remarkably, we find that {\it the first
order effect vanishes identically}, while the second order perturbation
contains singular terms rendering the equation ill-defined.  A concluding
section discusses a heuristic estimate of the quantum group parameter
$\kappa^{-1}$ for the $\kappa$-Dirac-Coulomb problem.

Confusingly, Dirac's quantum number in the Dirac-Coulomb problem is also called
kappa, and to avoid confusion, we shall henceforth denote the {\it inverse} of
the quantum group kappa parameter by $\epsilon$ and Dirac's quantum number, as
customary, by $\kappa$.

\section{The $\kappa$-Poincar\'{e}-Dirac-Coulomb Equation}

\indent \indent The $\kappa$-Poincar\'{e} quantum group has been developed by
Lukierski, Nowicki and Ruegg$^{8}$ and extended by the Lodz group,$^{10}$
and by Nowicki, Sorace and Tarlini.$^{9}$  We follow ref. 9 in
summarizing this structure.  The algebra structure is:
\begin{eqnarray}
&&\ [P_i, P_j] = 0\ , \hspace{2.26cm} [P_i, P_0]\ =\ 0\ , \nonumber \\
&&\ [M_i, P_j] = i \epsilon_{ijk} \; P_k\ , \hspace{1.06cm} [M_i, P_0]\ =\
0\ , \nonumber \\
&&\ [L_i, P_0] = i P_i\ , \hspace{1.95cm} [L_i, P_j]\ =\ i \epsilon^{-1}
\delta_{ij}   \, \sinh (\epsilon P_0) \ , \\  &&\ [M_i, M_j] = i \epsilon_{ijk}
\; M_{\kappa}\ , \quad\quad [M_i, L_j]\ =\ i \epsilon_{ijk} \; L_{\kappa} \ ,
\nonumber \\ &&\ [L_i, L_j] = -i \epsilon_{ijk} ( M_k \cosh (\epsilon P_0 )
-\frac{1}{4} \epsilon  \; P_{\kappa} \; P_l M_l )\ . \nonumber
\end{eqnarray}

Here $P_\mu \equiv \{P_{0}, P_{i}\}$ are the deformed generators for energy and
momenta, the $M_i$ are the spatial rotation generators (they close on an
undeformed Hopf co-commutative subalgebra), and the $L_i$ are the
deformed boost generators.

The coalgebra ($\Delta$ is the co-multiplication operation) and the antipode
($S$ is the quantum group analog of an inverse) are:
\begin{eqnarray}
\Delta M_i  =  M_i \otimes I + I \otimes M_i, \; \; \hspace{1.0in} \; \; \Delta
P_0 = P_0 \otimes I + I \otimes P_0, \nonumber \\
\Delta P_i  =  P_i \otimes {\mbox{exp}} (\frac{\epsilon P_0}{2}) +
{\mbox{exp}}  (- \frac{\epsilon P_0}{2}) \otimes P_i, \hspace{.5in}
\Delta L_i  =  L_i \otimes {\mbox{exp}} (\frac{\epsilon P_0}{2}) + \nonumber
\\
{\mbox{exp}} (- \frac{\epsilon P_0}{2}) \otimes L_i
 +  \frac{\epsilon}{2}\epsilon_{ijk}(P_i \otimes M_\kappa {\mbox{exp}}
( \frac{\epsilon P_0}{2}) + {\mbox{exp}} (- \frac{\epsilon P_0}{2})M_j
\otimes P_\kappa ) ,
\end{eqnarray}

\begin{equation}
S(P_\mu) = -P_\mu, \; S(M_i) = -M_i, \; S(L_i) = -L_i + \epsilon
\frac{3i}{2}P_i. \end{equation}

The deformed invariant operators of the $\kappa$-Poincar\'{e} group are:
\begin{eqnarray}
C_1  =  \left( \frac{2}{ \epsilon} \sinh ( \frac{\epsilon P_0}{2}) \right)^2
-P_i P_i, \; \; \  \
C_2  =  \left( \cosh ( \epsilon P_0) - \frac{\epsilon^{2} P_i
P_i}{4} \right) W_{0}^{2}-W_i W_i
\end{eqnarray}
where $W_0 = P_i M_i$ and $W_i = {\displaystyle \frac{1}{\epsilon}}
\sinh(\epsilon P_0) M_i + \epsilon_{ijk} P_j L_k$.

The Lorentz Lie algebra representation using the Dirac
$\gamma$-matrices is also a representation of the Poincar\'{e} Lie algebra with
the four-momentum generators represented by zero.  This representation also
fulfills the commutation relations (2.1) of the $\kappa$-Poincar\'{e} quantum
group.  (This is not surprising because in general the lowest dimensional
representation of a Lie algebra is a representation of the corresponding
quantum deformation.)  We can then ``add'' a spinless representation and the
Dirac $\gamma$-representation by using the co-multiplication of the
$\kappa$-Poincar\'{e} quantum group to obtain the desired spinorial
representation for $s= \frac{1}{2}$.  We obtain:
\begin{eqnarray}
{\cal P}_{\mu} = P_{\mu}, \hspace{1cm} {\cal M}_{i} = M_{i} + m_{i},
\hspace{1cm} {\cal L}_i = L_i + \mbox{exp} \left( -\epsilon \frac{P_0}{2}
\right) l_i - \frac{\epsilon}{2} \epsilon_{ i j k} m_j P_k,
\end{eqnarray}
where $m_i = \frac{i}{4} \epsilon_{ijk} \gamma_i \gamma_k \; \mbox{and}
\; l_i = - \frac{i}{2} \gamma_0 \gamma_i$.

The $\kappa$-deformed Dirac operator -- call it $\cal D$ -- is required to be
invariant under the generators $\{ {\cal P}_\mu , {\cal M}_i , {\cal L}_i \}$
in (2.5).  From the commutation relations (and properties of the Dirac
matrices) one can verify that the operator ${\cal D}$:
\begin{equation}
{\cal D} \equiv - {\mbox{exp}} ( - \frac{\epsilon P_0}{2} ) \gamma_{i} P_{i}
+ \gamma_{4} \frac{1}{\epsilon} \sinh (\epsilon P_{0}) - \frac{i \epsilon}{2}
\gamma_{4} P_{i} P_{i},
\end{equation} obeys the required relations:
$\left[ {\cal D}, {\cal L}_i \right] = \left[ {\cal D}, {\cal M}_i \right] =
\left[ {\cal D}, {\cal P}_\mu \right] = 0.$

Moreover, the square of $\cal D$ can be found to be:
\begin{equation}
{\cal D}^{2} = C_1 (1 + \frac{\epsilon^{2}}{4}  C_1) = - \frac{4}{3} C_2 .
\end{equation}

Thus the $\kappa$-Dirac equation may be written in the explicit
form:
\begin{equation}
\left( - {\mbox{exp}} ( - \frac{\epsilon P_0}{2 } ) \gamma_{i} P_{i} +
\gamma_{4} \frac{1}{\epsilon} \sinh (\epsilon P_{0}) - \frac{i\epsilon}{2}
\gamma_{4} P_{i} P_{i} \right) \psi = m (1 + \frac{\epsilon^{2} m^{2}}{4}
)^{\frac{1}{2}} \psi, \end{equation}
where $m^2 = C_1$.  (Note that in the limit $\epsilon \rightarrow 0$, one
recovers the usual Dirac equation.)

Let us expand the $\kappa$-Dirac equation, (2.8) (multiplied
for convenience by $ \textstyle {\mbox{exp}} \left( \frac{\epsilon P_{0}}{2}
\right) $ on both sides), in powers of $\epsilon$.  The resulting
equation, to an error $\approx \epsilon^{2}$, is: \begin{equation}
[(\gamma_4 P_0 - \gamma_i P_i) +  \frac{\epsilon}{2}
(\gamma_4 ( P_{0}^{2} - P_i P_i ) - m P_0)] \psi = m \psi.
\end{equation}

In order to obtain the $\kappa$-Dirac-Coulomb equation, it is necessary now to
gauge eq. (2.9), thereby introducing the Coulomb potential.  In
general, gauging of quantum groups is a current research
problem,$^{11}$ but for the $U(1)$ group
of electromagnetism and a one-body equation (no co-multiplication) the
straightforward gauging $P_\mu \rightarrow P_\mu - e A_\mu$ is quite
satisfactory.$^{12}$ (Note that in eq.(2.9),
unlike eq.(2.8), there are no operator ordering problems after the gauging that
would require symmetrization.)

Accordingly, we gauge eq. (2.9), $P_0 \rightarrow H + \displaystyle
{\frac{\alpha Z}{r}}$, introducing the (attractive) Coulomb potential.
We obtain
the first order $\kappa$-Dirac Coulomb equation:
\begin{equation}
H \psi = \{ ( \gamma_4 \gamma_i P_i + m \gamma_4 - \frac{\alpha Z}{r} ) +
\frac{\epsilon}{2} ( ( H + \frac{\alpha Z}{r} )^{2} - P_i P_i -
\gamma_4 m ( H + \frac{\alpha Z}{r} ) ) \} \psi.
\end{equation}

To an error of order $\epsilon^2$, we can identify $H$ in the perturbation
terms with the Dirac Hamiltonian ($H_{\cal D}$, eq.2.10 with $\epsilon = 0$ ).
The perturbation conserves angular momentum and parity.

Since the unperturbed (bound-state) Dirac-Coulomb problem has solutions
$(H_{\cal D} \rightarrow E_{\cal D}, \; N =$ principal quantum number,
$\kappa$ = Dirac's quantum number) degenerate in the sign of $\kappa$
(except for $\kappa = -N$), one would normally apply $2 \times 2$ degenerate
perturbation theory for an order $\epsilon$ calculation of the shifted energy
levels.  However, parity is conserved by the perturbation, so that the
$2 \times 2$ matrix becomes diagonal.  The diagonal matrix elements of
the perturbation in eq. (2.10) are to be taken between eigenstates of
the unperturbed Dirac-Coulomb Hamiltonian.  Using the quadratic Dirac
Hamiltonian for $P_i P_i$ we can put the
perturbation in a more convenient form.  Denoting the perturbation in
eq. (2.10) (the term multiplied by $\frac{\epsilon}{2}$) by $H_{pert}$
we find:  (Dirac's notation)
\begin{eqnarray}
H_{pert} = m^2 - \frac{i \alpha Z}{r^2} \gamma_4
{\mbox{\boldmath $\gamma \cdot$}} \hat{\mbox{\boldmath $r$}} - E_{\cal D} m
\gamma_4 - m \gamma_4 \frac{\alpha Z}{r}.
\end{eqnarray}

To our initial surprise, matrix elements of eq.(2.11) were found to be {\it
zero}.  In fact, the perturbation is {\it identically zero}.  It is easy to
show this: with (2.11) in operator form,
\begin{eqnarray}
H_{pert} = m^2 - \frac{1}{2}\{H_{\cal D}, m \gamma_4 \} - m\gamma_4
\frac{\alpha Z}{r}.
\end{eqnarray}

Using the Dirac Hamiltonian eq. (2.12) is seen to vanish.  {\it The first
order correction to the $\kappa$-Poincar\'{e} Dirac-Coulomb problem is
identically zero.}$^{13}$  This is quite remarkable and indicates that the
$\kappa$-Poincar\'{e} quantum group tends to make minimal changes in standard
structures.

\section{Further Remarks and Conclusion}

\indent \indent In an attempt to determine some experimental limit on the
$\kappa$-Poincar\'{e} length scale, it would be natural now to proceed to the
second-order corrections in the $\kappa$-Poincar\'{e} Dirac-Coulomb problem.
It is easily seen from eq.(2.8) that this procedure must fail, {\it since the
second-order perturbation is singular}.  This singularity comes from the term
$\frac{1}{6} \left(\frac{\alpha Z}{r} \right)^3$ introduced via
$\frac{1}{\epsilon} \sinh (\epsilon P_0)$ in (2.8), and cannot be eliminated.
(The full second-order correction is complicated and need not be given
here.)

Thus we are at an impasse in this attempt to bound the $\kappa$-Poincar\'{e}
length scale experimentally.  We remark, however, that there is a rather
speculative way to proceed further.  One of the motivations underlying the
quantum group approach is that the quantum parameter may possibly serve as a
convergence factor.  The singular terms in the $\kappa$-Poincar\'{e}-Dirac-
Coulomb problem enter through exponentials of the $P_0$ operator, which
suggests that finite (time) displacements occur.  Thus it is not unreasonable
to hope that the $\kappa$-Poincar\'{e} length scale may somehow cut-off the
singular terms in the Coulomb problem.  For the singular second order
perturbation term the cut-off enters logarithmically, and is thus relatively
insensitive to the cut-off.  Using the ground state second-order (cut-off)
energy shift determines, in fact, a length-scale that is self-consistent (the
cut-off scale in the singular integral is of the same order of magnitude as
that scale determined by the cut-off energy shifts).

The ground-state second-order (cut-off) energy shift is found to have the
order of magnitude:

\begin{eqnarray}
\Delta E (1s_{\frac{1}{2}}) \approx (m \epsilon)^2 \cdot (m \alpha^6).
\end{eqnarray}

Let us now compare this estimated shift, $\Delta E(1s_{\frac{1}{2}})$, with the
observed experimental limits$^{14}$ on the accuracy with which the
$1s_{\frac{1}{2}}$ eigen-energy is known.  A reasonable estimate from the
results in ref. 14 is that any deviation from the eigen-energy of the
$1s_{\frac{1}{2}}$ state is less than approximately $10^{-3}MHz$.  Since the
Rydberg, $\frac{1}{2} m \alpha^2$, is 3289 $THz$ this implies:
\begin{eqnarray}
(m \epsilon )^2 \left( m \alpha^6 \right) \approx  \left( m
\alpha^2 \right) \times 10^{-12}, \; \; \mbox{or} \; \; m
\epsilon \approx 10^{-2}.
\end{eqnarray}

This implies that the $\kappa$-Poincar\'{e} length scale is:
$\epsilon \approx 10^{-13} cm$ or smaller.  This self-consistent
estimate of the
$\kappa$-Poincar\'{e} length scale is both speculative and not very accurate,
despite the high experimental precision, in direct consequence of the
remarkable
properties found for the $\kappa$-Poincar\'{e} Dirac-Coulomb problem which has
no first-order perturbation and a singular second-order perturbative effect.

The referee has called our attention to the fact that the hydrogen
atom is a two--body problem and accordingly the required
co--multiplication of momenta must be considered. For the hydrogen
atom the relevant internal energy scale is $m_e$ which is small
($\approx 1/2000$) compared to the proton mass, so that the proton
motion is essentially non--relativistic. For this special case, going
to the center--of--momentum frame yields a first order quantum group
correction linear in the total momentum. Thus in the rest frame the
first order correction vanishes, validating our work. It is important
to recognize, however, that in the positronium system (where the two
components have equal mass) this argument is not justified and first
order co--multiplicative corrections may not be ruled out.
We wish to thank the referee for pointing out that the work of
Friedberg and Lee$^{15}$ discretizing time also gave no first order
corrections.

\section{Acknowledgement}

\indent \indent We would like to thank Professor Mark Raizen,
University of Texas (Austin) for discussions on precision measurements
in atomic hydrogen.

\vspace{.5in}

\noindent \large
{\bf REFERENCES AND FOOTNOTES} \normalsize

\begin{enumerate}
\item V.G. Drinfeld, {\it Proceedings of the ICM}, Berkeley, CA, edited by
A.M. Gleason (American Mathematical Society, Providence, RI, 1986), p. 798.

\item S. Woronowicz, Publ. RIMS (Kyoto Univ.) {\bf 23}, 117 (1987);
Commun. Math. Phys. {\bf 122}, 125 (1989).

\item L. Faddeev, N. Yu. Reshetikhin, and L. Takhtajan, in {\it Braid Group,
Knot Theory and Statistical Mechanics}, edited by C.N. Yang and M.L. Ge (World
Scientific, Singapore, 1989).

\item A.N. Kirillov and N. Yu. Reshetikhin, LOMI preprint E-9-88 (1988).

\item L.C. Biedenharn and M. Tarlini, Lett. Math. Phys. {\bf 20}, 271 (1990).

\item L.A. Takhtajan, Adv. Stud. Pure Math. {\bf 19}, 435 (1989).

\item E. Celeghini, R. Giachetti, E. Sorace and M. Tarlini, J. Math. Phys. {\bf
31}, 2548 (1990); {\bf 32}, 1155, 1159 (1991); ``Contraction of quantum
groups'' in Lect. Notes in Math. n. 1510, pg. 221 (Springer-Verlag, Berlin
1992).

\item J. Lukierski, A. Nowicki and H. Ruegg, Phys. Lett. {\bf B 293} (1992)
344.

\item A. Nowicki, E. Sorace and M. Tarlini, ``The Quantum Deformed Dirac
Equation from the $\kappa$-Poincar\'{e} Algebra'', preprint DFF 177/12/92
(Firenze, Italy), Phys. Lett. {\bf B} (in press).

\item S. Giller, J. Kunz, P. Kosinski, M. Majewski and P. Maslanka, ``{\it On
q-covariant Wave Functions}'', Lodz University Preprint, August 1992.

\item L. Castellani, ``Gauge Theories of Quantum Groups'', preprint DFTT-19/92
(Torino, Italy).

\item We wish to thank Leonardo Castellani for discussion on this point.

\item In a preprint just received, J. Lukierski, H. Ruegg and W. R\"{u}hl,
``From
$\kappa$-Poincar\'{e} Algebra to $\kappa$-Lorentz Quasigroup:  A Deformation of
Relativistic Symmetry'', preprint KL-TH-92/22 find that the first order
correction vanishes in the {\it non-relativistic} approximation.

\item M. Weitz, F. Schmidt-Kaler, and T.W. H\"{a}nsch, Phys.Rev.Lett. {\bf 68},
1120 (1992).

\item R. Friedberg and T.D. Lee, Nucl. Phys. {\bf B 225},1 (1983).
\end{enumerate}
\end{document}